\documentclass[11pt]{article}
\usepackage[english]{babel}
\usepackage{amsmath,amsthm}
\usepackage{amsfonts}
\usepackage{graphicx}
\usepackage{amsmath}
\usepackage{amsfonts}
\usepackage{rotating}
\usepackage{geometry}
\usepackage{booktabs}
\usepackage{color}
\usepackage{setspace}
\usepackage[dvipsnames]{xcolor}
\usepackage{pdflscape}
\usepackage[normalem]{ulem}
\def\Cov{\text{Cov}}
\def\Var{\text{Var}}
\newenvironment{sistema}{\left\lbrace\begin{array}{@{}l@{}}}{\end{array}\right.}

\theoremstyle{definition}

\theoremstyle{remark}

\numberwithin{equation}{section}
\begin{document}

\title{Stochastic Dividend Discount Model: A formula for the covariance of random stock prices}
\author{Arianna Agosto\footnote{Banca Carige SpA - Via Cassa di Risparmio 15, 16123 Genova, Italy} \and Alessandra Mainini\footnote{Dipartimento di Discipline matematiche, Finanza matematica ed Econometria - Universit\`a
Cattolica del Sacro Cuore - via Necchi 9, 20123, Milano, Italy} \and Enrico Moretto\footnote{corresponding author: \texttt{enrico.moretto@uninsubria.it} - Dipartimento di Economia, Universit\`a degli Studi dell'Insubria, via Monte Generoso 71, 21100 Varese, Italy, and CNR-IMATI, via A. Corti 12, 20133 Milano, Italy}}%
%\address{}%
%\thanks{}%
\date{}%
% ----------------------------------------------------------------
%\begin{abstract}
%
%\end{abstract}
\maketitle
% ----------------------------------------------------------------
\section*{Abstract}
Dividend discount models have been the first and most used methodology to determine the price of common stocks that, under a financial point of view,  is the sum of all discounted future dividends. The classical Gordon and Shapiro model (1956) deals with a deterministic setting in which the dividends' growth rate is constant.
Later, Hurley and Johnson (1994 and 1998) and Yao (1997) have introduced randomness assuming the growth rate to be described by a finite-state random variable. This leads to a two-fold consequence: a closed-form expression for the expected value of stock prices and a sequence of random dividends behaving in a Markovian fashion.
However, as expected values only provide a limited explanation of random phenomena, higher-order moments are needed.
This motivates the present contribution that provides an explicit formula for the covariance between stock prices when their random growth rates are (possibly) correlated. This formula has a number of applications such as the choice of the welfare maximizing portfolio in the standard portfolio selection model. The theoretical result is eventually applied to real market data. \\ \vspace{0.15 cm}
\textbf{Keywords}: Stock valuation, \ Dividend Discount Model, \ Markov chain, \ Financial risk

\section{Introduction}
\label{sec:1}

Dividend discount models (DDM in the following) have been the first attempt to find a financially correct pricing formula for common stocks. DDM date back to seminal works by Williams (1938) and Gordon and Shapiro (1956) where, in an entirely deterministic setting, the price of a common stock is obtained by discounting all future dividends per share, by means of a rate reflecting the company's riskiness. A first notable attempt to relax the confining assumption of a non-stochastic framework is due to Hurley and Johnson (1994, 1998), and Yao (1997). These authors introduce randomness, paving the road to Stochastic DDM (SDDM for short) by assuming that dividends' growth rate from one period to the next is described by a finite-state random variable. This choice leads to a Markovian sequence of non-stationary dividends, as found in Agosto and Moretto (2015). As a further improvement, Hurley (2013) assumes a continuous random variable for the growth rates.  All such contributions end up with an explicit formula for the expected value of the stock price. Recently, more complex settings have been studied: for instance, D'Amico (2013, 2016) generalizes SDDM by introducing a discrete-time semi-Markov chain.

However, it is well known that, in order to tackle a vast range of practical financial problems such as, for instance, portfolio selection and risk management, location measures are not enough. For a more reliable analysis, dispersion measures are also needed, as done in the Markowitz's portfolio selection model (1952), in which risk and performance  are measured in terms of, respectively, variances and means. A first answer to this issue in the SDDM framework has been provided by Agosto and Moretto (2015), who propose a closed-form expression for the variance of random stock prices.

As the expression of the portfolio's variance (\textit{i.e.}, a linear combination of stocks with given weights) requires covariances, to fill the remaining gap, this paper analytically derives  a formula for the covariance between random stock prices with (possibly) correlated random growth rates. A relevant result is that such covariance depends, in a monotonic way, on the covariance between growth rates; if growth rates are uncorrelated then random stock prices have zero covariance.

The structure of paper is the following: Section 2 summarizes previous results in SDDM while Section 3 illustrates the formula for the covariance between two stock prices. Section 4 contains an analysis using real market data; Section 5 eventually concludes.

\section{Review of SDDM}
\label{sec:2}
DDM provides an expression for the stock price if dividends per share grow at the constant geometric rate $g > - 1$ so that such dividends at time $j = 1, 2, \ldots$, is $d_j = d_0(1 + g)^j$ being $d_0$ the current one. If dividends are paid forever, and the discount rate $k > g$ is also constant, the value of the stock at time $0$ is
\[
P_0 = \sum_{j = 1}^{+\infty}\,\frac{d_j}{(1 + k)^j} = \frac{d_0(1 + g)}{k - g}.
\]
As said in Section \ref{sec:1}, this framework can be extended to encompass random dividends $\tilde d_j$. For instance Hurley and Johnson (1998) let the growth rate be the finite-state random variable
\begin{equation}
\tilde g =
\left\{
\begin{array}{rcccc}
\text{growth rates}  & \  & g_1,   & \ldots, & g_n \\
\text{probabilities} & \  & p_1,  & \ldots, & p_n,
\end{array}
\right.
\label{distribuzione}
\end{equation}
where $- 1 < g_1 < \cdots < g_n$, $p_i = \mathbb P\left[ \tilde g = g_i\right] > 0$, $i = 1, \ldots, n$, and $\sum_{i = 1}^n\,p_i = 1$. The sequence of these random dividends is Markovian, since it satisfies the recursive equation $\tilde d_{j + 1} = \tilde d_j\left( 1 + \tilde g\right)$. The random variable describing the current stock price is
\begin{equation}
\tilde P_0 = \sum_{j = 1}^{+\infty}\,\frac{\tilde d_j}{( 1 + k )^j}.
\label{random_P}
\end{equation}
If further $k > \bar g = \sum_{i = 1}^n\,g_ip_i$ then the expected value of $\tilde P_0$ is
\begin{equation}
\bar P_0
=
\frac{d_0( 1 + \bar g)}{k - \bar g}.
\label{HJ_Formula}
\end{equation}
Agosto and Moretto (2015) derive the variance of $\tilde P_0$:
\begin{equation}
\Var\left[\tilde P_0\right]
=
\frac{\Var\left[\tilde g\right]}{(1 + k)^2 - (1 + \bar g)^2 - \Var\left[\tilde g\right]}
\times\frac{(1 + k)^2}{(1 + \bar g)^2}\times\bar P_0^2.
\label{Var_P}
\end{equation}
$\Var\left[\tilde P_0\right]$  exists and is non-negative if $\Var\left[\tilde g\right] < (1 + k )^2 - \left(1 + \bar g\right)^2$. Note that it exists a range for $k$, $\bar g$, and $\Var\left[\tilde g\right]$, in which the expected stock price converges while variance does not.
In the following section we generalize this formula by calculating the covariance between two stock prices.

\section{Enlarging the scene: a formula for the covariance between stock prices}
\label{sec:3}
Consider two companies, say $A$ and $B$, with random growth rates $\tilde g_m$, $m = A,B$, that follow the joint distribution
\[
\pi_{cd} = \mathbb P\left[\left(\tilde g_A = g_{Ac}\right) \cap \left(\tilde g_B = g_{Bd}\right)\right] \geq 0
\;\;\;\;\;\;
\text{with}
\;\;\;\;\;\;
\sum_{c = 1}^n\,\sum_{d = 1}^n\,\pi_{cd} = 1,
\]
for $c,d = 1,2,\ldots,n$. The growth rates of the two companies are represented by the following random variables:
\begin{equation}
\tilde g_m
=
\left\{
\begin{array}{rcccc}
\text{growth rates}  & \  & g_{m1}, & \ldots,  & g_{mn} \\
\text{probabilities} & \  & p_{m1}, & \ldots,  & p_{mn},
\end{array}
\right.
\end{equation}
being, as in (\ref{distribuzione}), $- 1 < g_{m1} < g_ {m2} < \ldots < g_ {mn}$, and $\bar g_m  = \sum_{i  = 1}^n\,g_{mi}p_{mi}$. Assume also that $k_m > \bar g_m$, being $k_m $ the discount rate for company $m = A, B$. According to (\ref{random_P}) and (\ref{HJ_Formula}),
\[
\tilde P_{m0} = \sum_{j = 1}^{+ \infty}\,\frac{\tilde d_{mj}}{(1 + k_m)^j}
\;\;\;\;\;\;\;\;\;\;
\text{and}
\;\;\;\;\;\;\;\;\;\;
\bar P_{m0} =
\frac{d_{m0}( 1 + \bar g_m)}{k_m - \bar g_m},
\]
being $d_{m0}$ their current dividend. In order to obtain the covariance between $\tilde P_{A0}$ and $\tilde P_{B0}$, expected value $\mathbb E\left[\tilde P_{A0}\tilde P_{B0}\right]$ is needed. Under proper conditions,
\begin{equation}
\mathbb E\left[\tilde P_{A0}\tilde P_{B0}\right]
=
\sum_{j = 1}^{+ \infty}\,\sum_{p = 1}^{+ \infty}\,\frac{\mathbb E\left[\tilde d_{Aj}\tilde d_{Bp}\right]}{(1 + k_A)^j(1 + k_B)^p}.
\label{valore-atteso-prodotto}
\end{equation}
To determine $\mathbb E\left[\tilde d_{Aj}\tilde d_{Bp}\right]$, two cases need to be considered, namely, $j \leq p$ and $j > p$. If $j \leq p$ let
\[
d\left(s_{11},\,s_{12},\,s_{13},\,\ldots,\,s_{nn}\right)
=
d_{A0}d_{B0} \prod_{c,d = 1}^n\,\left(1 + g_{Ac}\right)^{s_{cd}}\left(1 + g_{Bd}\right)^{s_{cd}}
\]
denote the possible outcomes of time $j$ product $\tilde d_{Aj} \tilde d_{Bj}$,  when $\tilde d_{Aj}$ has grown  in $j$ steps  $s_{cd}$ times at rate $g_{Ac}$ while, at the same time,  $\tilde d_{Bj}$ has grown  in $j$ steps  $s_{cd}$ times at rate $g_{Bd}$, with $\sum_{c,d = 1}^n\,s_{cd} = j$. Probabilities of such outcomes are
\begin{equation}
\binom{j}{s_{11},s_{12},s_{13}\ldots,s_{nn}} \pi_{cd}^{s_{cd}}.
\end{equation}
On the other hand, let
\[
z_d = \sum_{c = 1}^n\,s_{cd},
\]
be the number of times in which dividend $ \tilde d_{Bj}$ has grown at rate $g_{Bd}$ in the first $j$ steps, regardless of the behavior of $ \tilde d_{Aj}$. If, overall,  $ \tilde d_{Bp}$ grows $w_d$ times at rate $g_{Bd}$, from $j + 1$ to $p$ it can grow at the same rate at most $r_d = \max\{w_d - z_d,0\}$ times. This allows to define a further random variable $\tilde d_{Bjp}$ that represents the behavior of $B$'s dividend between $j + 1$ and $p$. Possible outcomes of $\tilde d_{Bjp}$ are
\[
d\left(r_1,\ldots,r_n\right) = \prod_{d = 1}^n\, \left(1 + g_{Bd}\right)^{r_d},
\]
with $\sum_{d = 1}^n\,r_d = p - j$. The corresponding probabilities are
\[
\binom{p - j}{r_1,\ldots,r_n} p_{Bd}^{r_d}.
\]
The Markovian structure of the dividends' sequence leads to
\begin{align}
\bigskip
\mathbb E\left[\tilde d_{Aj}\tilde d_{Bp}\right]
& =
d_{A0}d_{B0} \times
\underbrace{\left(\sum_{s_{11} + \ldots + s_{nn} = j}\,
\prod_{c,d = 1}^n\,(1 + g_{Ac})^{s_{cd}}(1 + g_{Bd})^{s_{cd}}
\binom{j}{s_{11},\ldots,s_{nn}} \pi_{cd}^{s_{cd}}\right)}_{{(\ast)}} \notag \\
\bigskip
&
\hspace{1.6cm}\times
\underbrace{\left(\sum_{r_1 + \ldots + r_n = p - j}\,\prod_{d = 1}^n\,(1 + g_{Bd})^{r_d}\binom{p - j}{r_1,\ldots,r_n}p_{Bd}^{r_d}\right)}_{{(\ast\ast)}}.
\label{val-att-prod-div}
\end{align}
The other case, $j > p$, is analogous to the previous one if two changes are done. The first regards sum $(\ast)$ in (\ref{val-att-prod-div}), where $p$ replaces $j$ in $s_{11} + \ldots + s_{nn} = j$. The second change relates to sum ($\ast\ast)$ in  (\ref{val-att-prod-div}), where $g_{Bd}$ and $p_{Bd}$ are replaced by $g_{Ac}$ and $p_{Ac}$, and $r_d$ is replaced by $r_c$, which is defined analogously to $r_d$. Applying the multinomial theorem to (\ref{val-att-prod-div}) yields
\begin{align*}
\bigskip
\mathbb E\left[\tilde d_{Aj}\tilde d_{Bp}\right]
& =
d_{A0}d_{B0}
\left(\sum_{c = 1}^n\,\sum_{d = 1}^n\,(1 + g_{Ac})(1 + g_{Bd})\pi_{cd}\right)^j
\left(\sum_{d = 1}^n\,(1 + g_{Bd})p_{Bd}\right)^{p - j} \\
& = d_{A0}d_{B0}
\left[(1 + \bar g_A)(1 + \bar g_B) + \Cov[\tilde g_A,\tilde g_B]\right]^j\left(1 + \bar g_B\right)^{p - j},
\end{align*}
while, if $j > p$, then
\begin{align*}
\bigskip
\mathbb E\left[\tilde d_{Aj}\tilde d_{Bp}\right]
& =
d_{A0}d_{B0}
\left(\sum_{c = 1}^n\,\sum_{d = 1}^n\,(1 + g_{Ac})(1 + g_{Bd})\pi_{cd}\right)^p
\left(\sum_{c = 1}^n\,(1 + g_{Ac})p_{Ac}\right)^{j - p} \\
& = d_{A0}d_{B0}
\left[(1 + \bar g_A)(1 + \bar g_B) + \Cov[\tilde g_A,\tilde g_B]\right]^p\left(1 + \bar g_A\right)^{j - p}.
\end{align*}
If
\begin{equation}
G_m = \bar g_m + \frac{\Cov\left[\tilde g_m,\tilde g_l\right]}{1 + \bar g_l},\;\;\;m,l  = A,B,\;m\neq l,\label{capital-G}
\end{equation}
is seen as a risk-adjusted growth rate, then the following expression
\[
(1 + G_m)(1 + \bar g_l) = (1 + \bar g_A)(1 + \bar g_B) + \Cov\left[\tilde g_A,\tilde g_B\right],\;\;\;m,l  = A,B,\;m\neq l,
\]
holds, yielding
\begin{equation}
\mathbb E\left[\tilde d_{Aj}\tilde d_{Bp}\right]  %j=s, p=t
=
\begin{sistema}
\smallskip
d_{A0}d_{B0}\left(1 + G_A\right)^j\left(1 + \bar g_B\right)^p\; \; j \leq p \\
d_{A0}d_{B0}\left(1 + G_B\right)^p\left(1 + \bar g_A\right)^j \; \; j > p.
\end{sistema}
\label{prod-div}
\end{equation}
Finally, substituting (\ref{prod-div}) in (\ref{valore-atteso-prodotto}), leads to simple (but tedious) calculations that can be found in  Appendix \ref{sec:6}. At last, the covariance between stock prices $\tilde P_{A0}$ and $\tilde P_{B0}$ is
\begin{align}
\Cov\left[\tilde P_{A0},\tilde P_{B0}\right]
= &
\frac{\Cov\left[\tilde g_A,\tilde g_B\right]}{\displaystyle{\prod_{m = A,B}\,(1 + k_m) - \prod_{m = A,B}\,(1 + \bar g_m)} - \Cov\left[\tilde g_A,\tilde g_B\right]}
\times\prod_{m = A,B}\,\frac{1 + k_m}{1 + \bar g_m}\times\bar P_{m0},
\label{covarianza}
\end{align}
once the condition
\begin{equation}
\left\vert \prod_{i = A,B}\,(1 + \bar g_m) + \Cov\left[\tilde g_A,\tilde g_B\right]\right\vert < \prod_{m = A,B}\,(1 + k_m)
\label{cov-existence}
\end{equation}
is satified (see Appendix \ref{sec:6} for more details). Clearly, when $A  = B$ formula (\ref{Var_P}) is recovered. Note that condition (\ref{cov-existence}) ensures that the sign of the covariance between stock prices is the same of the covariance between growth rates, and is zero if dividends' growth rates are not correlated. Moreover, $\Cov\left[\tilde P_{A0},\tilde P_{B0}\right]$ monotonically increases in $\Cov\left[\tilde g_A,\tilde g_B\right]$, while it diverges whenever $\Cov\left[\tilde g_A,\tilde g_B\right]$ approaches $\left(1 + k_A\right)\left(1 + k_B\right) - \left(1 + \bar g_A\right)\left(1 + \bar g_B\right)$. This formula turns out to be useful in many issues; one of the most important is described in the following section.

\section{Covariance in practice: optimal portfolio choice}
\label{sec:4}
The main goal of portfolio theory is to identify the ``best'' investment strategy in terms of the efficient combination of stocks that maximizes some utility function. Here only two stocks are considered, say $A$ and $B$, whose random prices at time $1$ are $\tilde P_{A1}$ and $\tilde P_{B1}$. Their random returns are
\[
\tilde r_m = \frac{\tilde P_{m1} - \bar P_{m0}}{\bar P_{m0}},\;\;\;m = A,\,B.
\]
The random one-period return of portfolio $\mathbf x = [x_A,x_B]$ that invests in $A$ and $B$ is
\[
\tilde r(\mathbf x) = \tilde r_Ax_A  +\tilde r_Bx_B.
\]
A way to deal with the optimal trade-off between portfolio's risk and expected return is to evaluate a risky position by means of some risk-averse utility function. If the quadratic function $u(x) = x - 0.5\alpha x^2$,  $\alpha > 0$, $x \leq 1/\alpha$, is chosen, only the first two moments of $\tilde r_1$ are needed. In fact,
$$
\mathbb E\left[u\left(\tilde r(\mathbf x)\right) \right]
= \mathbb E\left[ \tilde r(\mathbf x) - 0.5\alpha \tilde r^2(\mathbf x)\right]
 = \bar r(\mathbf x) - 0.5\alpha \mathbb E\left[\tilde r^2(\mathbf x)\right],
$$
so that, as $\Var\left(\tilde r(\mathbf x)\right) = \mathbb E\left[\tilde r^2(\mathbf x)\right] - \mathbb E^2\left[\tilde r(\mathbf x)\right]$,
\[
\mathbb E\left[u\left( \tilde r(\mathbf x)\right) \right]
= \bar r(\mathbf x) - 0.5\alpha \left(\Var\left(\tilde r(\mathbf x)\right) + \bar r^2(\mathbf x)\right),
\]
where $\bar r(\mathbf x) = \bar r_Ax_A + \bar r_Bx_B$ and
\[
\Var\left(\tilde r(\mathbf x)\right) = \Var\left[\tilde r_A\right]x^2_A + 2\Cov\left[\tilde r_A,\tilde r_B\right]x_A x_B + \Var\left[\tilde r_B\right]x^2_B.
\]
The ``optimal'' portfolio is, then, the unique vector $\mathbf x^{\ast}$ that solves the constrained maximization problem
\[
\max_{\mathbf x \in\mathcal S}\,\mathbb E\left[ u\left( \tilde r(\mathbf x)\right) \right],
\]
where $\mathcal S = \{\mathbf x\in\mathbb R^2:\,x_A + x_B = 1$\}. Standard Lagrangian method yields
\begin{equation}
\label{x_A^*}
x_A^*
=
\frac{\alpha^{-1}\left(\bar r_A - \bar r_B\right) - \Cov\left[\tilde r_A,\tilde r_B\right]
+ \Var\left[\tilde r_B\right] - \bar r_B\left( \bar r_A - \bar r_B\right)}
{\left(\Var\left[\tilde r_A\right] - 2\Cov\left[\tilde r_A,\tilde r_B\right] + \Var\left[\tilde r_B\right]\right) + \left(\bar r_A - \bar r_B\right)}
\end{equation}
and assures that $\mathbf x^{\ast}$ is indeed a constrained maximizer for
$\mathbb E\left[u\left(\tilde r\right)\right]$. The crucial point in expressing $\tilde r_m$ is representing $\tilde P_{m1}$. Formula (\ref{random_P}) leads to
\[
\tilde P_{m1}
= (1 + k_m)\tilde P_{m0} - \tilde d_{m1},\;\;\;m = A,B,
\]
so that
\begin{equation*}
\bar P_{m1} = (1 + k_m)\bar P_{m0} -  d_{m0}\left(1 + \bar g_m\right),\\
\end{equation*}
\begin{equation*}
\Var\left[\tilde P_{m1}\right] = (1 + k_m)^2\Var\left[\tilde P_{m0}\right] - 2(1 + k_m)\Cov\left[\tilde d_{m1},\tilde P_{m0}\right] + d_{m0}^2\Var\left[\tilde g_m\right],
\end{equation*}
\begin{equation*}
\begin{split}
 \Cov\left[\tilde P_{A1},\tilde P_{B1}\right] = &
\Cov\left[\tilde P_{A0},\tilde P_{B0}\right]\prod_{m = A,B}(1 + k_m) - \sum_{m,l= A,B}^{m \neq l}(1 + k_m)\Cov\left[\tilde d_{m1},\tilde P_{l0}\right] \\
   & + d_{A0}d_{B0}\Cov\left[\tilde g_A,\tilde g_B\right].
\end{split}
\end{equation*}
To accomplish the task, covariances between single dividends and stock prices are to be determined. Using the fact that, again for each $m,l = A,B$ and $p \geq1$ (see (\ref{prod-div})),
\[
\mathbb E\left[\tilde d_{m1}\tilde d_{lp}\right]
=
d_{A0}d_{B0} \left(1 + G_m\right)\left(1 + \bar g_l\right)^p,
\]
it results that
\begin{equation*}
\begin{split}
\Cov\left[\tilde d_{m1},\tilde P_{l0}\right] = d_{m0}\frac{\Cov\left[\tilde g_m,\tilde g_l\right]}{1 + \bar g_l}\bar P_{l0};
\end{split}
\end{equation*}
in particular, for $m = l$,
\[
\Cov\left[\tilde d_{m1},\tilde P_{m0}\right] = d_{m0}\frac{\Var\left[\tilde g_m\right]}{1 + \bar g_m}\bar P_{m0}.
\]
These latter expressions eventually carry all the elements needed to determine (\ref{x_A^*}) and  the optimal portfolio strategy.

\subsection{An econometric analysis}

In order to empirically test the theoretical result presented in the previous sections, real market data (source: Bloomberg) of E.ON ($A$) and Saint-Gobain ($B$) are exploited. These two companies are, as of December 31, 2016, included in the Eurostoxx 50 market index, which encompasses the fifty European companies with the largest capitalization. The reason for choosing these companies relies on the fact that the historical series of their dividends is the amplest available, with data ranging from 1989 to 2016 for a total of 28 dividends. Further, their data satisfy the condition for which denominator of (\ref{Var_P}) is strictly positive, leading to positive variances for both stock prices. Table \ref{Tabella1} summarizes such data. It displays dividends paid in 1989 and 2016, the minimum and maximum yearly dividends growth rate in the time series as well as the geometric mean and  median of growth rates.

Discount rates $k$ for both companies are obtained by using the standard Capital Asset Pricing Model (CAPM) approach. For each company a linear regression model is estimated where the dependent variable is the excess stock return over the risk-free rate while the regressor (i.e. the market portfolio return of the CAPM) is the Eurostoxx 50 index return. In particular, in order to estimate the slope of the regression lines, that is Sharpe's $\beta $'s, weekly returns from January 2012 to December 2016 (for a total of 261 observations) are used. A proxy for the risk-less rate is $R_F = 0.5\%$. This choice is justified by the fact that, on one hand, the EUREPO rate index has been discontinued at the beginning of 2015 while, on the other, European interest rates have reached very low levels in the last years. As a proxy for the expected market return $R_M$, the logarithmic yearly mean of the Eurostoxx 50 index (i.e. $R_M = 6.905\%$) from January 2012 to December 2016 is adopted. The use of a smaller time window than the one considered for dividends allows the estimated discount rates not to be influenced by values very distant in time. The choice of a five-year period seems, under this point of view, adequate. According to the standard CAPM model the risk-adjusted discount rates are
\[
k_i = R_F + \beta_i\left(R_M - R_F\right),
\]
therefore  $k_A = 6.631\%$ and $ k_B = 7.943\%$.

The results of regressions are shown in Tables \ref{Tabella2a} and \ref{Tabella2b}, where parameter estimates, t-statistics and p-values are displayed. The slope parameter $\beta$ is significant in both models and its value is consistent with the common interpretation of CAPM. Estimated $\beta$ is indeed larger than $1$ for Saint-Gobain, belonging to the manufacturing sector which is believed to ``amplify'' stock market movements, while E.ON stocks show a slightly reduced sensitivity to systematic variability ($\beta < 1$), as it is commonly found for utilities sector companies. The constant parameter $\alpha $ turns out not to be significant in either models. As a measure of goodness-of-fit, we also calculate the R-squared value, which is satisfactory in both cases.

\begin{table}
\caption{Summary of dividends (and their rates of growth) times series: 1989-2016}
\label{Tabella1}
\centering
\begin{tabular}{l|cccccc}
\hline
Company       & $d_{1989}$ & $d_{2016}$    &Min       & Max        & Geom. Mean & Median    \\
\hline
E.ON          & $0.293$    & $0.5$          &$-0.4545$ & $0.2239$    & $0.02$     & $0.09091$ \\
Saint-Gobain  & $0.664$    & $1.24$          &$-0.4631$ & $0.25$      & $0.0234$   & $0.0303$  \\
\hline
\end{tabular}
\end{table}

\begin{table}
\caption{Linear regression output - Eurostoxx 50 vs E.ON: $R^2=0.3741$}
\label{Tabella2a}
\centering
\begin{tabular}{c|ccc}
\hline
Parameter & Estimate  & $t-$stat  & $P-$value               \\
\hline
$\alpha$  & $-0.0032$ & $-1.6591$ & $0.0983$                \\
$\beta$   & $0.9571$  & $12.4647$ & $2.8504 \cdot 10^{-28}$ \\
\hline
\end{tabular}
\end{table}

\begin{table}
\caption{Linear regression output - Eurostoxx 50 vs Saint-Gobain: $R^2=0.5639$}
\label{Tabella2b}
\centering
\begin{tabular}{c|ccc}
\hline
Parameter & Estimate  & $t-$stat  & $P-$value               \\
\hline
$\alpha$  & $0.0007$   & $0.4341$  & $0.6646$                \\
$\beta$   & $1.1621$  & $18.3010$  & $1.3869 \cdot 10^{-48}$ \\
\hline
\end{tabular}
\end{table}

The limited number of available dividends suggest\textbf{s} to consider, for both companies, only two alternative outcomes (i.e $g_1$ and $g_2$) for the dividends' rate of growth. Such values have been obtained in two different ways:
\begin{enumerate}
\item $g_1$ and $g_2$ are the geometric mean of the growth rates that happen to be below or above the geometric mean (as reported in Table \ref{Tabella1}), and
\item $g_1$ and $g_2$ are the median values (i.e. the first and third quartiles) of growth rates that are below or above their median (see Table \ref{Tabella1}).
\end{enumerate}
The reason for this choice is an attempt to investigate the effect of outliers in the rate of growth of dividends as it is well known that quartiles are descriptive statistics not influenced by extreme values.

Historical probabilities $\pi_{cd}$ are the ratios between the number of years in which the dividends' rates of growth have jointly been either below or above one of the two thresholds  above and the number of observations. Tables \ref{Tabella3a} and \ref{Tabella3b} report joint and marginal probabilities, along with the values for growth rates, obtained in the two ways just described. Each two-way table leads to different values for covariance between rates of growth. In the first case (geometric mean), covariance and correlation are, respectively, $0.043\%$ and $18.232\%$; such values become $0.036\%$ and $12.179\%$ when medians are considered. The reduction in correlation observed in the second case can be, at least partially, explained by the fact that dividends' rates of growth have reached, in 2007 and 2008, the abnormal values of $21.818\%$ and $22.388\%$ for E.ON and $25\%$ and $20.588\%$ for Saint-Gobain. Similar occurrences cannot be found elsewhere.

\begin{table}
\caption{E.ON vs Saint-Gobain joint probabilities' two-way table - geometric mean case}
\label{Tabella3a}
\centering
\begin{tabular}{c|cc|c}
\hline
& $g_{B1} = - 2.627\%$ & $g_{B2} = 5.100\%$ &   \\
\hline
$g_{A1} = - 5.019\%$ & $\pi_{11} = 0.25926$       &
$\pi_{12} = 0.18519$      & $p_{A1} = 0.44444$ \\
$g_{A2} = 7.390\%$  & $\pi_{21} = 0.22222$   & $\pi_{22} = 0.33333$  & $p_{A2} = 0.55556$ \\
\hline
& $p_{B1} = 0.48148$  & $p_{B2} = 0.51852$  &  \\
\hline
\end{tabular}
\end{table}

\begin{table}
\caption{E.ON vs Saint-Gobain joint probabilities' two-way table - median case}
\label{Tabella3b}
\centering
\begin{tabular}{c|cc|c}
\hline
 & $g_{B1} = 0.000\%$   & $g_{B2} = 8.688\%$ &  \\
 \hline
$g_{A1} = 0.000\%$      & $\pi_{11} = 0.28$  & $\pi_{12} = 0.24$  & $p_{A1} = 0.52$ \\
$g_{A2} =  13.810\%$ & $\pi_{21} = 0.2$   & $\pi_{22} = 0.28$  & $p_{A2} = 0.48$ \\
\hline
 & $p_{B1} = 0.48$  & $p_{B2}  = 0.52$ &                     \\
\hline
\end{tabular}
\end{table}

As $\Var\left[\tilde g_A\right] = 0.02431$ and $\Var\left[\tilde g_B\right] = 0.01447$, plugging available data into formul\ae \ (\ref{HJ_Formula}) and (\ref{Var_P}) yield the following values: $\bar P_{A0} = 11.01$ and $\bar P_{B0} = 22.65$, $\Var\left[\tilde P_{A0}\right] = 44.60$ and $\Var\left[\tilde P_{B0}\right] = 79.86$. Finally, plugging all data in (\ref{covarianza}), the covariance between stock prices of the two companies are $1.11872$ when geometric average growth rates are considered and $0.93951$ if, instead, medians are used.

These data are eventually used to perform the mean-variance analysis introduced at the beginning of this Section. For the sake of simplicity, only the geometric mean case is considered. The vector of expected returns and the matrix of variances and covariances are, respectively,
\[
\begin{pmatrix}
\bar r_A \\ \bar r_B
\end{pmatrix}
=
\begin{pmatrix}
2\% \\ 2.34\%
\end{pmatrix}
\;\;\;\;\;\;\;\;\;\;
\text{and}
\;\;\;\;\;\;\;\;\;\;
\begin{pmatrix}
\Var\left[\tilde r_A\right]            & \Cov\left[\tilde r_A,\tilde r_B\right]\\
\Cov\left[\tilde r_A,\tilde r_B\right] & \Var\left[\tilde r_B\right]
\end{pmatrix}
=
\begin{pmatrix}
0.4155 & 0.0051\\
0.0051 & 0.1798
\end{pmatrix}.
\]
\begin{figure}
\caption{Optimal investment in E.ON against $\alpha$.}
\label{figura}
\centering
\includegraphics[width=8cm]{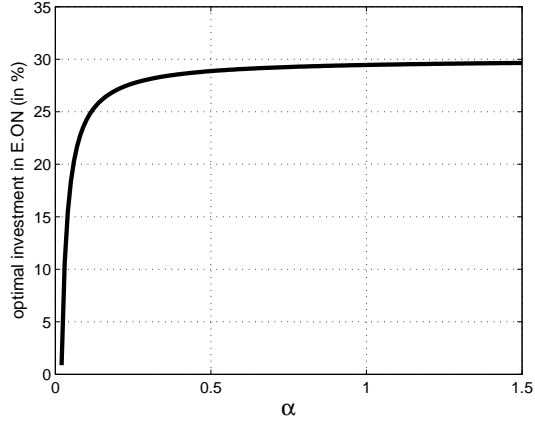}
\end{figure}
The minimum variance portfolio $\mathbf x_\text{min}$ is obtained minimizing $\Var\left[\tilde r (\mathbf x)\right]$ subject to the budget constraint. It results that $\mathbf{x}_\text{min} = \left[29.85\%,70.15\%\right]$, with $\bar r_\text{min} = 2.24\%$ and $\Var\left[\tilde r_\text{min}\right] = 0.1276$. According to (\ref{x_A^*}) the composition of portfolio $\mathbf x^{\ast}$ depends on the parameter $\alpha$: figure (\ref{figura}) shows the monotonic relationship between risk aversion and the quantity invested in E.ON. As $\alpha$ grows, risk aversion increases so that the fraction of wealth invested in E.ON increases until it reaches its minimum portfolio's level $\mathbf{x}_\text{min}$. This occurs because even if Saint-Gobain dominates, in terms of mean-variance, E.ON (that is: $\bar{r}_A < \bar{r}_B$ and, simultaneously, $\Var\left[\tilde r_A\right] > \Var\left[\tilde r_B\right]$), a combination of the two stocks (due to a covariance very close to $0$) can reduce risk an investor should bear in picking only one of the  asset. A proper portion of E.ON stock into the portfolio leads to diversification.

\section{Concluding remarks}
\label{sec:5}
In this article a closed-form formula for determining the covariance between stock prices that behave according to the SDDM is presented. This formula shows, as one expects, that the covariance between stock prices is strictly connected to the covariance of  their rates of growth. This means that the behavior of the covariance between random stock price is driven by the joint probabilities of their rates of growth. The formul\ae \ for the variance of SDDM stock prices and their covariances can be intended as a complete set of tools capable of investigating and likely extending existing results in corporate finance (e.g. Larson and Gonedes (1969) and Yagil (1987) analysis of exchange ratio determination in merging agreements) as well as financial mathematics (Value-at-Risk and Expected Shortfall risk measures for portfolios containing stocks whose prices are assumed to behave according to SDDM). On top of this, the fact that the formula for the covariance of stock prices is grounded on joint probabilities gives room for the introduction of copul\ae \ that, loosely speaking, allow to model a vast range of (not necessarily linear) links between the marginal behavior of the two stocks.

\section{Appendix}
\label{sec:6}

The analytic expression for
\[
\mathbb E\left[ \tilde P_{A0}\tilde P_{B0}\right]
=
\sum_{j = 1}^{+\infty}\,\sum_{p = 1}^{+\infty }\,
\frac{\mathbb E\left[ \tilde d_{Aj}\tilde d_{Bp}\right]}{(1 + k_A)^j(1 + k_B)^p}
\]
is obtained recalling (\ref{prod-div}). It results that
\begin{equation}
\mathbb E\left[ \tilde P_{A0}\tilde P_{B0}\right]
=
d_{A0}d_{B0}\sum_{p = 1}^{+\infty}\,
\left(
\underset{\text{\#1}}{\underbrace{\sum_{j = 1}^p\,\frac{(1 + G_A)^j(1 + \bar g_B)^p}{(1 + k_A)^j(1 + k_B)^p}}}
+
\underset{\text{\#2}}{\underbrace{\sum_{j = p + 1}^{+\infty}\,\frac{(1 + \bar g_A)^j(1 + G_{B})^{p}}{(1 + k_A)^j(1 + k_B)^p}}}\right).
\label{TheSomma}
\end{equation}
For ease of notation, let, for $i = A,B$,
\[
\gamma _{g_i} = \frac{1 + \bar g_i}{1 + k_i}
\hspace{1cm}
\text{and}
\hspace{1cm}
\gamma _{G_i} = \frac{1 + G_i}{1 + k_i}.
\]
Sum $\#1$ becomes
\[
\gamma _{g_B}^p\,\sum_{j = 1}^p\gamma _{G_A}^j
=
\frac{\gamma _{G_A}}{1 - \gamma _{G_A}}\left(1 - \gamma _{G_A}^p\right)\gamma_{g_B}^p,
\]
while sum $\#2$, that converges to a positive value as soon as $\bar g_A < k_A$, reduces to
\[
\gamma _{G_B}^p\sum_{j = p + 1}^{+\infty}\,\gamma _{g_A}^j
=
\frac{\gamma _{g_A}}{ 1-\gamma _{g_A}}\left(\gamma _{G_B}\gamma _{g_A}\right)^p.
\]
Observing that
\[
\gamma _{G_B}\gamma _{g_A}
=
\frac{\left(1 + \bar g_A\right)\left(1 + \bar g_B\right) + \Cov\left[\tilde g_A,\tilde g_B\right]}{\left(1 + k_A\right)\left(1 + k_B\right)}
=
\gamma _{G_A}\gamma _{g_B},
\]
sum (\ref{TheSomma}) results being
\begin{align*}
\smallskip
\mathbb E \left[ \tilde P_{A0}\tilde P_{B0}\right]
& = d_{A0}d_{B0}\sum_{p = 1}^{+\infty}\,\left(\frac{\gamma _{G_A}}{1 - \gamma_{G_A}}\gamma _{g_B}^p - \frac{\gamma _{G_A}}{1 - \gamma _{G_A}}
\left(\gamma _{G_A}\gamma _{g_B}\right)^p + \frac{\gamma _{g_A}}{1 - \gamma_{g_A}}\left(\gamma _{G_B}\gamma _{g_A}\right)^p\right)\\
\smallskip
& = d_{A0}d_{B0}\left(\frac{\gamma _{G_A}}{1 - \gamma_{G_A}}\sum_{p = 1}^{+\infty }\,\gamma _{g_B}^p
+
\left(\frac{\gamma _{g_A}}{1 - \gamma_{g_A}} - \frac{\gamma _{G_A}}{1 - \gamma_{G_A}}\right)\sum_{p = 1}^{+\infty }\,\left(\gamma _{G_B}\gamma _{g_A}\right)^p\right).
\end{align*}
Now, $\sum_{p = 1}^{+\infty}\,\gamma_{g_B}^p$ converges to
\[
\frac{\gamma _{g_B}}{1 - \gamma _{g_B}} = \frac{1 + \bar g_B}{k_B - \bar g_B} > 0
\]
if $\bar g_B < k_B$, whereas $\sum_{p = 1}^{+\infty}\left( \gamma_{G_B}\gamma_{g_A}\right)^p$ converges to
\[
\frac{\gamma _{G_A}\gamma _{g_B}}{1 - \gamma _{G_A}\gamma _{g_B}}
=
\frac{\left(1 + \bar g_A\right)\left(1 + \bar g_B\right) + \Cov\left[\tilde g_A,\tilde g_B\right]}
{\left(1 + k_A\right) \left(1 + k_B\right) - \left(1 + \bar g_A\right)\left(1 + \bar g_B\right) -
\Cov\left[\tilde g_A,\tilde g_B\right]}
\]
if
\[
\vert\left(1 + \bar g_A\right)\left(1 + \bar g_B\right)+ \Cov\left[ \tilde g_A,\tilde g_B\right]\vert
< \left(1 + k_A\right)\left(1 + k_B\right).
\]
Moreover, observe that
\[
\bar P_{m0} = d_{m0}\frac{\gamma _{g_m}}{1 - \gamma_{g_m}},\hspace{1cm} m = A,B.
\]
Covariance between $\tilde P_{A0}$ and $\tilde P_{B0}$ is, then,
\begin{align}
\smallskip
& \Cov\left[ \tilde P_{A0}\tilde P_{B0}\right]\notag\\
\smallskip
& = d_{A0}d_{B0}\left(\frac{\gamma _{G_A}}{1 - \gamma_{G_A}}\frac{\gamma _{g_B}}{1 - \gamma _{g_B}}
+
\left(\frac{\gamma _{g_A}}{1 - \gamma_{g_A}} - \frac{\gamma _{G_A}}{1 - \gamma_{G_A}}\right)\frac{\gamma _{G_A}\gamma _{g_B}}{1 - \gamma _{G_A}\gamma _{g_B}}
- \frac{\gamma _{g_A}}{1 - \gamma_{g_A}}\frac{\gamma _{g_B}}{1 - \gamma_{g_B}}\right)\notag\\
\smallskip
& = d_{A0}d_{B0}\left(\frac{\gamma _{G_A}}{1 - \gamma_{G_A}}\frac{\gamma _{g_B}}{1 - \gamma _{g_B}}
+
\left(\frac{\gamma _{g_A}}{1 - \gamma_{g_A}} - \frac{\gamma _{G_A}}{1 - \gamma_{G_A}}\right)\frac{\gamma _{G_A}\gamma _{g_B}}{1 - \gamma _{G_A}\gamma _{g_B}}
- \frac{\gamma _{g_A}}{1 - \gamma_{g_A}}\frac{\gamma _{g_B}}{1 - \gamma_{g_B}}\right)\notag\\
\smallskip
& = d_{A0}d_{B0}\left(\frac{\gamma _{G_A}}{1 - \gamma_{G_A}} - \frac{\gamma _{g_A}}{1 - \gamma_{g_A}}\right)
\left(\frac{\gamma _{g_B}}{1 - \gamma _{g_B}}
-
\frac{\gamma _{G_A}\gamma _{g_B}}{1 - \gamma _{G_A}\gamma _{g_B}}\right)\notag\\
\smallskip
& = d_{A0}d_{B0}\times
\frac{\gamma _{G_A} - \gamma _{g_A}}{\left(1 - \gamma_{G_A}\right)\left(1 - \gamma_{g_A}\right)}\times
\frac{\gamma _{g_B}\left(1 - \gamma _{G_A}\right)}{\left(1 - \gamma _{g_B}\right)\left(1 - \gamma _{G_A}\gamma _{g_B}\right)}\notag\\
\smallskip
& = d_{A0}\frac{\gamma _{g_A}}{1 - \gamma_{g_A}}\times
d_{B0}\frac{\gamma _{g_B}}{1 - \gamma_{g_B}}\times
\frac{\gamma _{G_A} - \gamma _{g_A}}{\gamma _{g_A}\left(1 - \gamma _{G_A}\gamma _{g_B}\right)}\notag\\
\smallskip
& = \bar P_{A0}\bar P_{B0}
\times\frac{\gamma _{G_A} - \gamma _{g_A}}{\gamma _{g_A}\left(1 - \gamma _{G_A}\gamma _{g_B}\right)}.
\label{verso-la-covarianza}
\end{align}
Now, recall the definition of $\gamma _{G_A}$, $\gamma _{g_A}$, $\gamma _{g_B}$, and $G_A$. Then,
\[
\gamma _{G_A} - \gamma _{g_A}
=
\frac{\Cov\left[\tilde g_A,\tilde g_B\right]}{(1 + k_A)(1 + \bar g_B)},
\]
and
\[
\gamma _{g_A}\left(1 - \gamma _{G_A}\gamma _{g_B}\right)
=
\frac{(1 + \bar g_A)\left((1 + k_A)(1 + k_B) - (1 + \bar g_A)(1 + \bar g_B) - \Cov\left[\tilde g_A,\tilde g_B\right]\right)}
{(1 + k_A)^2(1 + k_B)}
\]
Substituting these expressions in (\ref{verso-la-covarianza}),
\begin{align*}
\smallskip
& \Cov\left[ \tilde P_{A0}\tilde P_{B0}\right]\\
\smallskip
& = \bar P_{A0}\bar P_{B0}\times
\frac{\Cov\left[\tilde g_A,\tilde g_B\right]}{(1 + k_A)(1 + \bar g_B)}
\times\frac{(1 + k_A)^2(1 + k_B)}{(1 + \bar g_A)\left((1 + k_A)(1 + k_B) - (1 + \bar g_A)(1 + \bar g_B) - \Cov\left[\tilde g_A,\tilde g_B\right]\right)}\\
\smallskip
& = \bar P_{A0}\bar P_{B0}\times
\frac{(1 + k_A)(1 + k_B)}{(1 + \bar g_A)(1 + \bar g_B)}
\times\frac{\Cov\left[\tilde g_A,\tilde g_B\right]}{(1 + k_A)(1 + k_B) - (1 + \bar g_A)(1 + \bar g_B) - \Cov\left[\tilde g_A,\tilde g_B\right]}.
\end{align*}
This proves the claimed formula.


\begin{thebibliography}{90}
\bibitem{AgoMor2015}
Agosto A., Moretto E.: Variance matters (in stochastic dividend discount models), Annals of Finance 11, 283-295 (2015).
\bibitem{Ami2013}
D'Amico G.: A semi-Markov approach to the stock valuation problem, Annals of Finance, 9(4), 589-610 (2013).
\bibitem{Ami2016}
D'Amico G.: Generalized semi-Markovian dividend discount model: risk and return. arXiv. http://arxiv.org/abs/1605.02472 (2016). Accessed 13 July 2016.
\bibitem{GoSha1956}
Gordon M.J., Shapiro E.: Capital equipment analysis: the required rate of profit, Management Science 3, 102-110 (1956).
\bibitem{Hur2013}
Hurley W.J.: Calculating firstmoments and confidence intervals for generalized stochastic dividend discount models, Journal of Mathematical Finance 3, 275-279 (2013).
\bibitem{HurJoh1994}
Hurley W.J., Johnson L.D.: A realistic dividend valuation model. Financial Analysts Journal, 50(4), 50-54 (1994).
\bibitem{HurJoh1998}
Hurley W.J., Johnson L.D.: Generalized Markov dividend discount models, Journal of Portfolio Management, 25(1), 27-31 (1998).
\bibitem{LarGon1969}
Larson K.D., Gonedes N.J.: Business combinations: an exchange ratio determination model, Accounting Review, 444, 720-728 (1969).
\bibitem{Wil1938}
Williams J.B.: The Theory of Investment Value. Harvard University Press, Cambridge  (1938).
\bibitem{Yagil1987}
Yagil J.: An exchange ratio determination model for mergers: a note, Financial Review, 22, 1, 195-202.
\bibitem{Yao1997}
Yao Y.F.: A trinomial dividend valuation model, Journal of Portfolio Management 23(4), 99-103 (1997).
\end{thebibliography}
\end{document}